\documentclass[oneside,reqno,12pt]{amsart} 

\usepackage{rangecite}
\usepackage{euscript}

\newcommand{\mathscript}{\EuScript}

\newcommand{\Lag}{\mathscript{L}}
\newcommand{\cR}{\mathscript{R}}
\newcommand{\vev}[1]{\left\langle #1 \right\rangle}

\numberwithin{equation}{section}

\makeatletter
    \def\serieslogo@{\vtop to 0pt{\noindent\scriptsize\ppn\parindent\z@}}
    \let\@setcopyright\@empty
\makeatother

\begin{document}

\def\ppn{HEP-TH/9602075, UPR-691T}
\title[SUSY Breaking Induced by Higher-Derivative Supergravitation]
   {Soft Supersymmetry Breaking Induced by\\
   Higher-Derivative Supergravitation in\\
   the Electroweak Standard Model}
\author{Ahmed Hindawi, Burt A. Ovrut, and Daniel Waldram}
\thanks{Published in Physics Letters B \textbf{381} (1996), 154--162}
\maketitle
\vspace*{-0.3in}
\begin{center}
\small{\textit{Department of Physics, University of Pennsylvania}} \\
\small{\textit{Philadelphia, PA 19104-6396, USA}}
\end{center}

\begin{abstract}

We show how spontaneous supersymmetry breaking in the vacuum state 
of higher-derivative supergravity is transmitted, as explicit soft 
supersymmetry-breaking terms, to the effective Lagrangian of the standard
electroweak model. The general structure of the soft supersymmetry
breaking terms is presented and a new scenario for understanding the 
gauge hierarchy problem, based on the functional form of these terms,
is discussed.

\vspace*{\baselineskip}


\end{abstract}

\renewcommand{\baselinestretch}{1.2} \large \normalsize

\vspace*{\baselineskip} 

\section{Introduction}

A definitive answer to the question of how supersymmetry is broken in 
phenomenologically relevant theories of particle physics, be they 
supergravitational theories or superstrings, remains elusive. One candidate for 
a mechanism of supersymmetry breaking is the Polonyi model and its many 
generalizations \cite{PP-KFKI-93,PLB-126-215,PRD-27-2359,PR-110-1}. This 
approach requires the explicit introduction of new chiral supermultiplets, in 
addition to those of the standard supersymmetric electroweak model and 
supergravity. Furthermore, the interactions of such fields with observable 
matter must be of a specific form; that is, they must form a ``hidden'' sector. 
Why such fields should exist, and how they come to be hidden from ordinary 
matter, is not well understood. A second candidate for a mechanism of 
supersymmetry breaking is non-perturbative, using gaugino condensation in a 
strongly interacting sector \cite{PLB-125-457,PR-162-169}. Again, this mechanism 
requires extra superfields in the theory, in this case strongly interacting 
vector multiplets, and this sector must be sufficiently hidden from observable 
matter. Generically, it is not well understood why such a mechanism should 
occur, although in superstring theories this approach seems better motivated. In 
this paper, we will present what we believe to be a very different mechanism for 
supersymmetry breaking, which seems to overcome at least some of the 
shortcomings of the afore-mentioned approaches.

We begin by considering only the fields of the standard model and the graviton. 
In the usual way, we will assume that these fields are actually members of four-
dimensional, $N=1$ supermultiplets: chiral multiplets for quarks, leptons and 
Higgs fields, vector multiplets for the strong and electroweak gauge fields, and 
the supergravitational multiplet for the graviton. These are the only 
superfields we will introduce, which distinguishes our approach from those 
discussed above. It is traditional, when generalizing the standard 
supersymmetric model to include supergravity, to admit higher-dimensional 
interaction terms composed of products of fields suppressed by the Planck mass, 
but to disallow higher-derivative interactions, be they purely gravitational or 
involving matter. Generically, there is no justification for this. Higher-
derivative terms will appear at the same order when supermatter is coupled to 
supergravity. It is often said that such terms may indeed be present, but are 
irrelevant at low momentum since they will be suppressed by powers of momenta 
over the Planck mass. We wish to stress that this statement is naive. The reason 
is the following. It is well known that, in non-supersymmetric theories, higher-
derivative terms in the equations of motion correspond to new degrees of 
freedom, in addition to the degrees of freedom of the original theory. This 
situation is amplified in supersymmetric theories where higher-derivative terms 
not only produce new bosonic and fermionic degrees of freedom, but also cause 
fields that were auxiliary, and, hence, not physical, in the original theory to 
become propagating \cite{NPB-138-430}. We want to re-emphasize at this point 
that one is not introducing new superfields, rather the new degrees of freedom 
are arising out of derivatives of fields, or auxiliary fields, that have already 
been introduced. In two recent papers \cite{NPB-471-409,PP-UPR-685T}, we have 
shown that these new degrees of freedom can have non-trivial potential energies 
possessing multiple vacua with vanishing cosmological constant. It is essential 
to realize that one must first identify the vacuum structure of these new 
degrees of freedom, and expand around one of the vacua, before taking the low-
momentum limit. While it is true that the fluctuations around the vacuum will 
decouple at low momenta, the effect of the vacuum itself will not. It 
generically couples to the usual matter fields and survives to low momentum, 
much as the vacuum state of a grand-unified theory determines the gauge group 
structure of the low-energy effective theory. It is in this sense that the above 
statement is naive. This understood, it is important to ask what effect a 
non-trivial vacuum state can have on the low-energy effective matter Lagrangian. 
In a previous paper, we have shown that there exist vacua of the new degrees of 
freedom of pure higher-derivative supergravitation that spontaneously break 
supersymmetry. In this paper, we will couple such theories to the supersymmetric 
standard model and show that, at low momentum, such a non-trivial vacuum 
produces soft supersymmetry-breaking terms of fundamental phenomenological 
importance. Indeed, this mechanism seems sufficient to properly account for 
supersymmetry breaking in the standard supersymmetric model.

There is one further issue that we wish to touch upon before presenting our 
results. It is well known that higher-derivative theories tend to have at least 
some extra degrees of freedom whose propagation behavior is ghost-like. This 
seems to always be the case for matter fields \cite{PR-79-145} and is frequently 
the case in the gravitational sector \cite{GRG-9-353,PRD-53-5583,PRD-53-5597}. A 
number of authors have pointed out that, at least in the context of gravity, 
this apparent ghost-like behavior may be illusory 
\cite{PLB-97-77,Haw-QFTQS,HEP-TH-9601082}. This is a complicated issue that we 
would like to avoid in this paper. We have shown in a recent publication 
\cite{PP-UPR-685T} that there is a consistent and interesting higher-derivative 
extension of supergravity that is, in fact, completely ghost-free. For 
simplicity, we will, in this paper, take this to be the higher-derivative 
extension of the standard model coupled to gravity. Hence, the theory discussed 
in this paper will be ghost-free. Why nature chooses this ghost-free extension 
of the standard model, or whether generic higher-derivative theories are 
actually consistent as well, will be left for future research.

\section{Higher-Derivative Supergravity}

In this section we will consider pure supergravitation only. Ordinary $N=1$ 
Einstein supergravity is described by the Lagrangian
\begin{equation}
   \mathscript{L} = - 3 M_P^2 \int d^4\theta E,
   \label{eq:one}
\end{equation}
where $E$ is the superdensity and $M_P$ the Planck mass. 
It is well known that the minimal supergravity 
multiplet contains the graviton ${e_m}^a$, the gravitino ${\psi_m}^\alpha$, a
complex scalar $M$ and a real vector $b_m$ as component fields. Expressed in
terms of these component fields, Lagrangian \eqref{eq:one} becomes
\cite{WB-SS}
\begin{equation}
  e^{-1}  \mathscript{L} = M_P^2 \left( - \tfrac{1}{2}\cR 
           - \tfrac{1}{3}MM^{\ast} + \tfrac{1}{3}b^{m}b_{m} \right)
       + \tfrac{1}{2}\epsilon^{klmn} \left(
           \bar{\psi}_{k}\bar{\sigma}_{l}{\tilde\mathscript{D}}_{m}\psi_{n} 
           - \psi_{k}\sigma_{l}{\tilde\mathscript{D}}_{m}\bar{\psi}_{n}\right),
\label{eq:two}
\end{equation}
which describes bosonic Einstein gravitation coupled to a gravitino. The $M$ and 
$b_m$ fields are non-propagating auxiliary fields which can be eliminated from 
the Lagrangian using their equations of motion $M=0$ and $b_m=0$. It is 
important to note that it is not supersymmetry that is forcing $M$ and $b_m$ to 
be auxiliary, but rather it is the explicit choice of the Lagrangian 
\eqref{eq:one}. This Lagrangian was chosen because it describes the simplest 
supersymmetric extension of ordinary bosonic Einstein gravitation. However, were 
one to change it to include, for instance, higher-derivative gravitational 
terms, then it is no longer necessary that $M$ and $b_{m}$ be auxiliary fields. 
That is, they might begin to propagate and become physical. This is indeed the 
case, as was first noted at the linearized level in \cite{NPB-138-430} and was 
established on the non-linear level in \cite{PLB-190-86}. The same phenomenon in 
new minimal supergravity was discussed in \cite{NPB-306-160,PLB-254-132}.

Recently \cite{PP-UPR-685T}, we constructed the most general class of ghost-free 
higher-derivative supergravity theories. The Lagrangian for such theories is 
given by
\begin{equation}
   \mathscript{L} = -3M_{P}^{2}\int{d^{4}\theta Ef \left(
      \frac Rm,\frac{R^\dag}{m} \right)},
\label{eq:three}
\end{equation}
where $f$ is an arbitrary dimensionless real function and we have chosen to 
normalize the scalar curvature superfield $R$ by an arbitrary mass $m$. 
Expressed in terms of the component fields, this Lagrangian becomes
\begin{multline}
 e^{-1} \mathscript{L} = M_{P}^{2}\Big\{
-\tfrac12 (f+Mf_M+M^*f_{M^*}-4MM^*f_{MM^*}- 2 b^mb_m f_{MM^*}) \mathscript{R} \\
- \tfrac{3}{4} f_{MM^*} \mathscript{R}^2
+ 3 f_{MM^*} \partial^m M \partial_m M^* - 3 f_{MM^*} 
(\nabla^m b_m)^2 + \cdots \Big\},
\label{eq:four}
\end{multline}
where $f_{MM^*} = \partial^2f/\partial M\partial M^*$ and we have dropped all 
fermionic terms and all bosonic terms inessential for this discussion. We see 
from this expression that Lagrangian \eqref{eq:three} describes the 
supersymmetric extension of the higher-derivative bosonic $\cR+\cR^2$ theory. 
Furthermore, we see explicitly that both the complex $M$ field and the 
longitudinal part of the $b_{m}$ field are now propagating physical fields, as 
anticipated. It is well known that, in addition to the helicity-two graviton, 
$\cR+\cR^2$ bosonic gravitation contains a new propagating scalar field. 
Therefore, Lagrangian \eqref{eq:three} contains four new bosonic scalar degrees 
of freedom in addition to the graviton. The fermionic superpartners of these new 
degrees of freedom can be identified in the higher-derivative fermionic terms 
that occur, but were not shown, in component Lagrangian \eqref{eq:four}. These 
new degrees of freedom must arrange themselves into two new chiral 
supermultiplets.

Written in this higher-derivative form, the physical content of this theory is 
obscure. However, it can be shown \cite{PLB-190-86,PP-UPR-685T} that by 
performing a super-Legendre transformation, Lagrangian \eqref{eq:three} can be 
written in an equivalent form which consists of two chiral supermultiplets, 
denoted by $\Phi$ and $\Lambda$, coupled to ordinary Einstein supergravity with 
a particular K\"ahler potential and superpotential. Specifically, we showed in 
\cite{PP-UPR-685T} that Lagrangian \eqref{eq:three} is equivalent to the 
Lagrangian
\begin{equation}
   \mathscript{L} = \int d^{4}\theta E \left\{ 
      - 3 M_{P}^{2} \exp\left(-\frac13\frac{K}{M_P^2}\right)
      + \frac{W}{2R} + \frac{W^{\dagger}}{2R^{\dagger}} \right\},
\label{eq:five}
\end{equation}
where
\begin{equation}
\begin{aligned}
   K &= - 3 M_P^2 \ln \left\{f\left(\frac{\Phi}{M_P},
           \frac{\Phi^\dag}{M_P}\right)
        +\frac{\Lambda+\Lambda^{\dagger}}{M_{P}} \right\}, \\
   W &= 6 m \Phi\Lambda. 
\end{aligned}
\label{eq:six}
\end{equation}
Written in this form, it is relatively straightforward to determine the vacuum 
structure of the theory. Using the well-known formula for the potential energy 
\cite{WB-SS}, we find that
\begin{equation}
   V=12 \left\{ f\left(\frac{A}{M_P},\frac{A^*}{M_P}\right)
        + \frac{B+B^{\ast}}{M_{P}} \right\}^{-2} U(A,B),
\label{eq:seven}
\end{equation}
where
\begin{equation}
   U=M_P^2|A|^2 \Big( f - 2f_A A - 2f_{A^{\ast}}A^{\ast}
          + 4f_{AA^{\ast}}|A|^2 \Big)
       - f_{AA^{\ast}}^{-1}\Big| B-M_{P}f_{A}A + 
          2M_{P}f_{AA^{\ast}}|A|^{2}\Big|^2
\label{eq:eight}
\end{equation}
and $A$ and $B$ are the lowest components of chiral superfields $\Phi$ and 
$\Lambda$ respectively.  There is typically a local minimum of the potential at 
$\vev{A}=\vev{B}=0$, with vanishing cosmological constant, where supersymmetry 
remains unbroken. We refer to this minimum as the trivial vacuum. Generically, 
however, we find that such potentials have other local minima. These non-trivial 
vacua can be arranged to have zero cosmological constant and, rather remarkably, 
are found to spontaneously break supersymmetry. It must be kept in mind that we 
are analyzing a theory of pure higher-derivative supergravitation. This theory 
involves the gravity supermultiplet only; no new fundamental supermultiplets 
have been introduced. What these results tell us is that the extra scalar 
degrees of freedom in such theories can have non-trivial vacua with vanishing 
cosmological constant which can spontaneously break supersymmetry. There is no 
need for a Polonyi-type field, gaugino condensates, or any other mechanism. This 
result seems sufficiently important that we will give a concrete example. Let us 
consider a function $f$ of the form
\begin{equation}
   f\left(\frac Rm,\frac{R^\dag}{m} \right)
       = 1 - 2\frac{RR^{\dagger}}{m^{2}} 
           + \frac{1}{9}\frac{(RR^{\dagger})^2}{m^4}.
\label{eq:nine}
\end{equation}
We find that the associated potential energy has precisely two local minima, 
each with vanishing cosmological constant. The first is the trivial minimum at 
$\vev{A}=\vev{B}=0$ which does not break supersymmetry. However, we find a 
second minimum, actually a ring of minima with zero cosmological constant, at
\begin{equation}
\begin{aligned}
   \vev{A} &= M_P e^{i\theta}, \\
   \vev{B} &= \tfrac43 M_P.
\label{eq:ten}
\end{aligned}
\end{equation}
Computing the K\"ahler covariant derivatives at this vacuum, we find that
\begin{equation}
\begin{aligned}
   \vev{D_A W} &= 32mM_P, \\
   \vev{D_B W} &= -\tfrac{15}{2} e^{i\theta} mM_P.
\end{aligned}
\label{eq:eleven}
\end{equation}
It follows that supersymmetry is spontaneously broken with strength 
$(mM_P)^{1/2}$. The associated gravitino mass is found to be
\begin{equation}
   m_{3/2}=\tfrac{27}{8}m.
\label{eq:twelve}
\end{equation}
We now proceed to determine the effect of such a non-trivial vacuum on the low-
energy effective Lagrangian of the standard model.

\section{Higher-Derivative Supergravity Coupled to Matter}

In this section, we will consider the matter superfields of the supersymmetric 
standard model and then couple these fields to higher-derivative supergravity. 
Since the couplings of the vector multiplets are essentially fixed by gauge 
invariance, we will here only consider the chiral supermultiplets. These we will 
generically label by $Y_i$, suppressing all gauge indices. The supersymmetric 
standard model Lagrangian can be written in terms of a flat K\"ahler potential 
and a specific superpotential $g(Y_i)$ as
\begin{equation}
\Lag_{\text{Matter}} = \int d^4\theta \sum_i Y_i Y_i^\dag 
+ \left\{ \int d^2\theta g(Y_i) + \text{h.c.} \right\}.
\label{matter-flat}
\end{equation}
The superpotential is trilinear in $Y_i$ except for a possible Higgs bilinear 
term. In the following analysis, however, we do not need to use the exact form 
of $g$. It suffices to assume that it leads to a vacuum state where all 
$\vev{Y_i}$ either vanish, or are of the order of the electroweak scale. We will 
also take $\vev{g}=0$ without loss of generality. Since supersymmetry is not 
observed at low-energy, the full model must also include explicit soft 
supersymmetry-breaking terms. The purpose of this section is to show that, when 
the standard model is coupled to higher-derivative supergravity, such soft terms 
are induced naturally when supersymmetry is spontaneously broken in the 
gravitational sector.

The supergravity extension of Lagrangian \eqref{matter-flat} is usually taken to 
be
\begin{equation}
   \Lag = \int d^4\theta E \left\{ 
      - 3 M_P^2 \exp\left(-\tfrac13 M_P^{-2} \sum_i Y_iY_i^\dag \right) 
      + \frac{g}{2R} + \frac{g^\dag}{2R^\dag} \right\}.
\label{matter-curved}
\end{equation}
In addition to Einstein supergravitation, Lagrangian \eqref{matter-curved} 
introduces a restricted set of higher-dimensional operators suppressed by powers 
of $M_P$. These operators involve products of the matter fields, but do not 
contain higher-derivative interactions. However, this is by no means the most 
general extension of Lagrangian \eqref{matter-flat} to the Planck scale. 
Generically, in addition to the Planck mass suppressed terms contained in 
\eqref{matter-curved}, one can add higher-derivative terms involving both matter 
superfields and supergravity. There is a general class of such theories, where 
the higher-derivative terms are associated with the supergravity chiral 
superfield $R$, which are ghost-free. In this paper, we will restrict the 
discussion to theories of this type. A natural ghost-free higher-derivative 
extension of Lagrangian \eqref{matter-flat} is of the form
\begin{equation}
   \Lag = \int d^4\theta E \left\{ 
      -3 M_P^2 F\left(\frac Rm,\frac{R^\dag}{m},
          \frac{Y_i Y_i^\dag}{M_P^2} \right)
      + \frac{g}{2R} + \frac{g^\dag}{2R^\dag} \right\},
\label{matter-hd2}
\end{equation}
where $F$ is a dimensionless real function and $m$ is some mass-scale which does 
not have to be related to the Planck mass. We will assume, for the time being, 
that $m$ is less than $M_P$ by at least a few orders of magnitude. Further, we 
will assume that all the coefficients in the function $F$ are of order unity, so 
that $m$ and $M_P$ are the only two mass scales in the theory.

Just as in the case of pure higher-derivative supergravity, we can make a super-
Legendre transformation to put Lagrangian \eqref{matter-hd2} into an equivalent 
second-order form. The transformed theory describes Einstein supergravity 
coupled to matter superfields $Y_i$ plus two extra chiral supermultiplets, 
$\Phi$ and $\Lambda$. The superfields $\Phi$ and $\Lambda$ represent the new 
degrees of freedom arising from the higher-derivative supergravitational terms. 
Making the super-Legendre transformation gives the following K\"ahler potential 
and superpotential
\begin{equation}
\begin{aligned}
   K & = - 3 M_P^2 \ln \left\{ 
      F\left( \frac{\Phi}{M_P}, \frac{\Phi^\dag}{M_P},
            \frac{Y_i Y_i^\dag}{M_P^2}\right) 
      + \frac{\Lambda+\Lambda^\dag}{M_P} \right\}, \\
   W & = 6 m \Phi \Lambda + g(Y_i).
\end{aligned}
\label{K&W}
\end{equation}
The associated scalar potential energy is then found to be
\begin{equation}
   V = 12 e^{2K/3M_P^2} \left(U_1 + U_2 + U_3\right),
\label{pe}
\end{equation}
where
\begin{equation}
\begin{aligned}
   U_1 &= m^2 AA^* \left( F - 2 A F_A -2 A^* F_{A^*} + 4AA^* F_{AA^*} 
\right), \\
   U_2 &= - \frac12 \frac{m}{M_P} \left( A^* g + A g^* \right)
       - m^2 P_i (F_{ij^*})^{-1} P_j^*, \\
   U_3 &= - \frac{m^2}{M_P^2} \frac{\det F_{ij^*}}{\det X} \Big| B - M_P (A F_A
       - 2 AA^* F_{AA^*} - P_i (F_{ij^*})^{-1} F_{Aj^*}) \Big|^2,
\end{aligned}
\label{potential}
\end{equation}
and we define
\begin{equation}
\begin{aligned}
   P_i &= A F_i - 2 AA^* F_{iA^*} - \frac{g_i}{6mM_P}, \\
   X  &= \begin{pmatrix} F_{AA^*} & F_{Aj^*} \cr F_{iA^*} & F_{ij^*} 
         \end{pmatrix}.
\end{aligned}
\end{equation}
As in the previous section, we write $A$ and $B$ for the lowest components for 
$\Phi$ and $\Lambda$ respectively, and introduce $y_i$ for the lowest component 
of $Y_i$, while $F_A = \partial F/\partial A$ and $F_i=\partial F/\partial y_i$. 
It is convenient to expand the function $F$ as
\begin{equation}
   F\left(\frac{\Phi}{M_P},\frac{\Phi^\dag}{M_P},
         \frac{Y_iY_i^\dag}{M_P^2} \right) 
      = f\left(\frac{\Phi}{M_P},\frac{\Phi^\dag}{M_P}\right) 
         + \sum_i c_i \left(\frac{\Phi}{M_P},\frac{\Phi^\dag}{M_P}\right) 
             \frac{Y_i Y_i^\dag}{M_P^2} 
         + \cdots
\label{exp}
\end{equation}
and insert this into potential energy \eqref{pe}. If we ignore, for the time 
being, all terms associated with  matter superfields $Y_i$, then the potential 
energy reduces to the pure higher-derivative supergravity potential for the 
function $f$ given in equation \eqref{eq:seven} of the previous section. There 
we demonstrated, for a wide class of functions $f$, that the potential energy 
has a stable, local minimum with non-vanishing vacuum expectation values 
$\vev{A}$ and $\vev{B}$ of the order of $M_P$. Furthermore, the cosmological 
constant can be chosen to be zero and, generically, supersymmetry is 
spontaneously broken with strength $(mM_P)^{1/2}$. We, henceforth, assume $f$ is 
chosen to have these properties. Now let us restore the matter fields $Y_i$. We 
find that potential energy \eqref{pe}, although complicated, still possesses a 
vacuum with zero cosmological constant in which $\vev{A}$ and $\vev{B}$ are 
determined only by $f$ and have the above properties. The $y_i$ fields either 
have $\vev{y_i}=0$, or non-zero values of the order of the electroweak scale, 
depending on the choice of the superpotential $g$.

We will now show that, at low-energy, the fluctuations around this vacuum 
reproduce the standard supersymmetric model described by Lagrangian 
\eqref{matter-flat} along with a specific set of soft  supersymmetry-breaking 
terms. In this paper, we will consider the bosonic part of the Lagrangian only. 
We write $A=\vev{A}+a$ and $B=\vev{B}+b$ but, since  $\vev{y_i}\ll M_P$, leave 
the $y_i$ fields unexpanded. Inserting these expressions into the Lagrangian 
associated with \eqref{K&W} and dropping all terms suppressed by $M_P^{-1}$ or 
more, we find that the low-energy theory separates into two pieces: a pure $y_i$ 
part and a separate hidden sector depending on fields $a$ and $b$, which have 
masses of order $m$, but do not interact with the $y_i$ fields. We will, 
henceforth, ignore this hidden sector since, with the possible exception of 
cosmological consequences, it is physically irrelevant. The low-energy effective 
Lagrangian for the matter fields $y_i$ is found to be
\begin{equation}
   \mathscript{L}_{\text{eff}} 
      = 3 e^{2\vev{K}/3M_P^2} \sum_i \vev{c_i} |\partial_m y_i|^2 
          - V_{\text{eff}}(y_i),
\end{equation}
where
\begin{equation}
   V_{\text{eff}}(y_i) = V_{\text{SUSY}}(y_i) + V_{\text{Soft}}(y_i).
\end{equation}
The term $V_{\text{SUSY}}$ is given by
\begin{equation}
   V_{\text{SUSY}}(y_i) = - \tfrac13 e^{2\vev{K}/3M_P^2} 
       \sum_i \frac{|g_i|^2}{\vev{c_i}}.  
\end{equation}
Provided that $\vev{c_i}<0$, the kinetic energy term and $V_{\text{SUSY}}$ can, 
after appropriate rescaling of the $y_i$ fields, be written in the form of 
Lagrangian \eqref{matter-flat} with a modified superpotential $g$. Hence, these 
terms are supersymmetric. The remaining part, $V_{\text{Soft}}$, of the low-
energy potential is found to be
\begin{equation}
   V_{\text{Soft}}(y_i) = \sum_i m_{1,i}^2 |y_i|^2 
       + \left\{ m_2 g + \sum_i m_{3,i} y_i g_i + \text{c.c.} \right\},
\label{VSoft}
\end{equation}
where
\begin{equation}
\begin{aligned}
   m_{1,i}^2 &= 48 \frac{m^2}{M_P^2} \vev{AA^*}^2 e^{2\vev{K}/3M_P^2} 
      \vev{\frac{\partial^2 c_i}{\partial A\partial A^*} 
      - c_i^{-1} \frac{\partial c_i}{\partial A} 
          \frac{\partial c_i}{\partial A^*}}, \\
   m_2 &= -6 \frac{m}{M_P} \vev{A^*} e^{2\vev{K}/3M_P^2}, \\
   m_{3,i} &= - 2 \frac{m}{M_P} \vev{A^*} e^{2\vev{K}/3M_P^2} 
      \left\{ 2 \vev{A\frac{\partial\ln c_i}{\partial A}} -1 \right\}.
\end{aligned}
\label{soft}
\end{equation}
Unlike the previous case, $V_{\text{Soft}}$ cannot be written in the form of 
Lagrangian \eqref{matter-flat} and, therefore, corresponds to genuine soft 
supersymmetry breaking. Expressions \eqref{VSoft} and \eqref{soft} are quite 
general, valid for a wide class of superpotentials $g$ subject to the minor 
constraints discussed previously. They simplify somewhat if we specialize to the 
case in which the matter superpotential is a homogeneous polynomial of degree 
$n$. We then have
\begin{equation}
\sum_i y_i g_i = n g.
\end{equation}
This is the case in the $\mathrm{R}$-parity invariant, supersymmetric standard 
model where the Higgs bilinear term is disallowed. Furthermore, since, in this 
model, the terms in $g$ are trilinear, it follows that $n=3$. Therefore, 
$V_{\text{Soft}}$ in \eqref{VSoft} simplifies to
\begin{equation}
   V_{\text{Soft}} = \sum_i m_{1,i}^2 |y_i|^2 
      + \left\{ \sum_i m'_{3,i} y_i g_i + \text{c.c.} \right\},
\label{n=3}
\end{equation}
where 
\begin{equation}
\begin{aligned}
   m_{1,i}^2 &= 48 \frac{m^2}{M_P^2} \vev{AA^*}^2 e^{2\vev{K}/3M_P^2} 
          \vev{\frac{\partial^2 c_i}{\partial A\partial A^*} 
       - c_i^{-1} \frac{\partial c_i}{\partial A} 
          \frac{\partial c_i}{\partial A^*}}, \\
   m'_{3,i} &= - 4 \frac{m}{M_P} e^{2\vev{K}/3M_P}
       \vev{AA^* \frac{\partial\ln c_i}{\partial A}}.
\end{aligned}
\label{n=3:m}
\end{equation}
We conclude that for a wide class of theories, including the supersymmetric 
standard model, spontaneous supersymmetry breaking in the higher-derivative 
supergravity sector is transmitted to the low-energy effective Lagrangian of 
observable matter as a specific set of soft, explicit supersymmetry-breaking 
operators. This result is reminiscent of the Polonyi mechanism with a general 
K\"ahler potential \cite{PLB-126-215,PRD-27-2359}, but, unlike that mechanism, 
it arises from pure supergravitation and does not require the introduction of a 
hidden supersymmetry-breaking sector.

The soft supersymmetry-breaking terms will lead, through radiative corrections, 
to the spontaneous breakdown of electroweak symmetry. If we assume, recalling 
$\vev{A}\simeq M_P$, that the $\vev{c_i}$-dependent coefficients of $m$ or $m^2$ 
in \eqref{soft} or \eqref{n=3:m} are of order unity then the scale of the 
electroweak symmetry breaking will be set by $m$. In this case, one must take 
$m\simeq 10^2 \text{GeV}$. The degree of fine-tuning required to do this is 
exactly the same as in the Polonyi models, and, hence, this is a completely 
viable approach to electroweak symmetry breaking. However, the higher-derivative 
supergravity theory described in this paper has a second, and quite novel, 
possibility for solving the hierarchy problem, in the context of the 
$\mathrm{R}$-parity invariant, supersymmetric standard model. Note that the 
coefficients of the soft supersymmetry-breaking terms in \eqref{n=3} are 
explicit combinations of the vacuum expectation values of functions of $c_i$. 
Above, we assumed that these coefficients were of order unity. This then 
necessitated taking $m$ to be of the order of the electroweak scale. Note, 
however, that if, for some reason, these combinations are small, of the order of 
$10^{-16}$ or so, then the parameter $m$ can be chosen to be large. Thus, 
although the supersymmetry breaking in the matter sector is of the order of the 
electroweak scale, the gravitino mass and the masses of the hidden sector fields 
would be, generically, large. Consequently, in this case the so-called gravitino 
and Polonyi cosmological problems \cite{PLB-131-59} are automatically solved. 
Very small values for the $\vev{c_i}$-dependent coefficients are not as 
unexpected or unnatural as they may first appear to be. For example, suppose 
that
\begin{equation}
   c_i \left( \frac{A}{M_P},\frac{A^*}{M_P} \right) = \text{constant},
\label{condition}
\end{equation}
for all values of $i$. It is immediately clear from \eqref{n=3} that 
$V_{\text{Soft}}$ vanishes for any value of $m$. This is true despite the fact 
that supersymmetry is broken with strength $(mM_P)^{1/2}$ in the supergravity 
sector. This breaking is simply not transmitted to the low-energy observable 
theory. Condition \eqref{condition} is very simple and it is intriguing to 
speculate whether it could arise naturally as a consequence of a symmetry of the 
higher-derivative theory. Now suppose the functions $c_i$ are not strictly 
constant, but, instead, are slowly varying. It follows that $V_{\text{Soft}}$ no 
longer vanishes. However, these terms can be very small, conceivably of 
electroweak strength, independent of the value of $m$. Such slowly varying 
functions could arise naturally, for example, from small, perhaps non-
perturbative, breaking of the symmetry that enforced condition 
\eqref{condition}. This mechanism opens the possibility of theories involving 
the Planck mass only, with the gauge hierarchy arising from the structure of the 
functions $c_i$ and not from the fine-tuning of the parameter $m$. The results 
of this section all involved approximations in which it was assumed that $m\ll 
M_P$. To explore this new mechanism, we will now consider the case where 
$m=M_P$.

\section{Coupling to Matter With $m=M_P$}

We now repeat the analysis of the last section, but here we take $m=M_P$. It 
follows that we can no longer drop terms proportional to $m/M_P$. As a result, 
we must keep higher-order terms in the expansion of $F$. That is, instead of the 
expansion \eqref{exp}, we now consider 
\begin{multline}
   F\left(\frac{\Phi}{M_P},\frac{\Phi^\dag}{M_P},
          \frac{Y_iY_i^\dag}{M_P^2}\right)
      = f\left(\frac{\Phi}{M_P},\frac{\Phi^\dag}{M_P}\right)
          + \sum_i c_i
                \left(\frac{\Phi}{M_P},\frac{\Phi^\dag}{M_P}\right)
                \frac{Y_iY_i^\dag}{M_P^2} \\
          + \sum_{ij} e_{ij}
                \left(\frac{\Phi}{M_P},\frac{\Phi^\dag}{M_P}\right)
                \frac{Y_iY_i^\dag}{M_P^2}\frac{Y_jY_j^\dag}{M_P^2}
          + \cdots.
\label{Fexpand}
\end{multline}
For simplicity, we will consider only the $\mathrm{R}$-parity invariant 
supersymmetric standard model with no Higgs bilinear term. Exactly as in the 
previous section, the theory has a vacuum state where $\vev{A}$ and $\vev{B}$ 
take the values set by the pure higher-derivative supergravity potential energy 
associated with the function $f$, while $\vev{y_i}$ are zero or of the order of 
the electroweak scale. Again, we assume that $f$ is such that the vacuum 
spontaneously breaks supersymmetry with zero cosmological constant. However the 
scale of this breaking is now $M_P$. Furthermore, the masses of the fluctuations 
in $A$ and $B$ around the vacuum are now generically also of the order of $M_P$ 
and, hence, they completely decouple from the low-energy effective Lagrangian. 
We find that the low-energy Lagrangian is given by 
\begin{equation}
   \mathscript L_{\text{eff}} = 
      3 e^{2\vev{K}/3M_P^2} \sum_i \vev{c_i} |\partial_my_i|^2 
      - V_{\text{eff}}(y_i),
\end{equation}
where
\begin{equation}
V_{\text{eff}} = V_{\text{SUSY}}(y_i) + V_{\text{Soft}}(y_i) 
                  + V_{\text{Quartic}}(y_i).
\end{equation}
The terms $V_{\text{SUSY}}$ and $V_{\text{Soft}}$ have exactly the same form as 
in the previous section, but now with $m=M_P$. What is new here is that the 
potential energy also contains potentially hard supersymmetry-breaking operators
of the form
\begin{equation}
   V_{\text{Quartic}}(y_i) = \sum_{ij} \lambda_{ij} |y_i|^2 |y_j|^2,  
\end{equation}
where 
\begin{multline}
   \lambda_{ij} = \frac{12\vev{AA^*}e^{2\vev{K}/3M_P^2}}{M_P^2} 
      \,\Bigg\langle \left| 1-2A \frac{\partial \ln c_ic_j}{\partial A} 
           \right|^2 e_{ij}  
      + 2\left[ 1-2A^*\frac{\partial \ln c_ic_j}{\partial A^*} \right] 
          A \frac{\partial e_{ij}}{\partial A} \\
      + 2\left[ 1-2A\frac{\partial \ln c_ic_j}{\partial A} \right] 
            A^* \frac{\partial e_{ij}}{\partial A^*} 
      + 4AA^* \frac{\partial^2 e_{ij}}{\partial A\partial A^*} \Bigg\rangle \\
      - \frac{e^{\vev{K}/3M_P^2}}{M_P^2} \Big( 
         \vev{c_i} m_{1,j}^2 + \vev{c_j} m_{1,i}^2 \Big)
      - \frac{e^{-2\vev{K}/3M_P^2}}
         {48M_P^2\vev{\left(AA^*\right)^2f_{AA^*}}} m_{1,i}^2 m_{1,j}^2.
\label{lambda-def}
\end{multline}
In fact, a quartic term of this type, albeit proportional to $m/M_P$, was 
present in the previous section, as it is in the Polonyi models. Since in these 
theories $m\ll M_P$, such terms are automatically suppressed and, hence, are not 
discussed. In the present case, however, $m=M_P$ and these quartic terms are not 
automatically suppressed. In this paper, it will suffice to point out that the 
$\lambda_{ij}$ coefficients can always be made to vanish or to be very small, 
for any choice of $c_i$, by appropriately adjusting the functions $e_{ij}$ in 
equation \eqref{Fexpand}. Once this is done, $V_{\text{Soft}}$ can be made to be 
of the order of the electroweak scale by choosing the functions $c_i$ to vary 
sufficiently slowly, exactly as was discussed in the previous section. To be 
more concrete, let us suppose that, for some reason, 
\begin{equation}
   c_i \left( \frac{A}{M_P},\frac{A^*}{M_P} \right) = \text{constant},
\label{r-condition}
\end{equation}
for all $i$ and 
\begin{equation}
   e_{ij} \left( \frac{A}{M_P},\frac{A^*}{M_P} \right) = 
       \frac{h_{ij}(A/M_P) + \left(h_{ij}(A/M_P) \right)^*}
           {\left|A/M_P\right|}, 
\label{t-condition}
\end{equation}
for all $i$ and $j$, where $h_{ij}$ are arbitrary functions of
$A$. Putting these expressions into \eqref{lambda-def} immediately
yields 
\begin{equation}
   \lambda_{ij} = 0, 
\end{equation}
for all $i$ and $j$. Therefore,the quartic terms exactly vanish. Furthermore, 
since $c_i=\text{constant}$, we also have $V_{\text{Soft}}=0$. In this case 
then, despite the fact that supersymmetry is spontaneously broken at order $M_P$ 
in the supergravity sector, none of this breaking is transmitted to the 
low-energy effective Lagrangian. If the conditions \eqref{r-condition} and 
\eqref{t-condition} are now slightly altered, then the quartic terms will remain 
suppressed whereas an acceptable gauge hierarchy will be generated by the 
non-vanishing but small $V_{\text{Soft}}$ terms. Conditions \eqref{r-condition} 
and \eqref{t-condition} are relatively simple and could arise naturally as the 
consequence of a symmetry of the higher-derivative theory. Small, perhaps 
non-perturbative, breaking of these symmetries would then give rise to the 
observed gauge hierarchy. This intriguing possibility will be discussed 
elsewhere \cite{HOW}. We want to emphasize again that since, in these theories, 
both the gravitino and the supersymmetry-breaking scalar field masses are of 
order $M_P$, the gravitino and Polonyi cosmological problems \cite{PLB-131-59} 
are solved. Furthermore, since the new degrees of freedom associated with 
higher-derivative supergravitation have masses of order $M_P$, this scenario 
could well arise within the context of superstrings \cite{PLB-388-512}.

Finally it is worth noting that the Lagrangian \eqref{matter-hd2} is not the 
most general form of a ghost-free higher-derivative theory of supergravity 
coupled to matter. In general, one would not separate off the superpotential 
term $g/2R$ but, rather, include it in the function  $F$, so that $F$ becomes a 
function of $Y_i$ and $Y^\dag_i$ separately, giving 
\begin{equation}
   \Lag = - 3 M_P^2 \int d^4\theta E F(R,R^\dag,Y_i,Y_i^\dag).
\end{equation}
As we will describe elsewhere \cite{HOW}, extending the analysis of this paper 
to the general case leads to essentially the same results, with only some minor 
modifications to the form of the low-energy effective theory.

\section*{Acknowledgments}

This work was supported in part by DOE Grant No.\ DE-FG02-95ER40893 and 
NATO Grand No.\ CRG-940784.

\end{document}